\documentclass[12pt]{article}
\usepackage{epsfig}
\usepackage{amssymb}
\usepackage{pslatex}

\makeatletter

\@addtoreset{equation}{section}
\makeatother
\newcommand{\bea}{\begin{eqnarray}}
\newcommand{\eea}{\end{eqnarray}}
\newcommand{\be}{\begin{eqnarray}}
\newcommand{\ee}{\end{eqnarray}}

\def\rmd{{\rm d}}

\begin{document}

\begin{titlepage}
\vskip1cm
\begin{flushright}
$\mathbb{UOSTP}$ {\tt 10101}
\end{flushright}
\vskip1.25cm
\centerline{\large \bf 
Thermal Aspects of ABJM theory: Currents and Condensations 
}
%\centerline{\Large \bf and} 
%\centerline{\Large \bf
%its  type IIA superstring dual} 
\vskip1.25cm
\centerline{  Dongsu Bak$^a$ and  Sangheon Yun$^b$
}
\vspace{1.25cm}
\centerline{\sl a) Physics Department, University of Seoul, Seoul
130-743 {\rm Korea}}
\vskip0.25cm
\centerline{\sl b) Korea Institute for Advanced Study,
Hoegiro 87, Dondaemun-gu, Seoul 151-747 {\rm Korea}}
\vskip0.25cm
\centerline{\tt dsbak@uos.ac.kr,  sanhan@kias.re.kr }
\vspace{1.5cm}
\centerline{ABSTRACT}
\vspace{0.75cm}
\noindent
To study thermal aspects of the ABJM theory in the strongly coupled
regime, we carry out the $\mathbb{CP}_3$ invariant dimensional reduction
of the type IIA supergravity down to four dimensions. We then investigate
zero and finite temperature responses of the operators which are dual
to the AdS scalar and vector fields. Two scalar operators are shown to 
have finite-temperature condensations by coupling of constant 
source term.  The currents dual to the massless and massive gauge 
fields are not induced by  coupling of constant boundary 
vector potential, which implies that the phase described by 
black brane background is not superconducting. We also discuss 
a generalization to  charged (dyonic) black holes.

\vspace{1.75cm}
%\centerline{\today}
\end{titlepage}
%%%%%%%%%%%%%%%%%%%%%%%%%%%%%%%%%%%%%%%%%%%%%%%%%%%%%%%%%%%%%%%

\section{Introduction}
There has been remarkable progress in understanding of
 the AdS/CFT correspondence between string 
theories and conformal field theories\cite{Maldacena:1997re}. 
The type IIB string theory
on AdS$_5\times S^5$ is dual to the four dimensional ${\cal N}=4$ super
Yang-Mills theory\cite{Maldacena:1997re}, 
which is by now tested to a convincing 
level.  
Thus one may now apply this duality to explore some new physics or 
to gain  some precise understanding of strongly coupled
regime of field theories where  our direct field-theoretic
understanding is limited. % in general. 
The Janus
correspondence as a  
controlled deformation of AdS$_5$/CFT$_4$\cite{Bak:2003jk}
is one such example, which leads to  
interesting 
predictions for the interface conformal field 
theories\cite{Clark:2004sb}. 
It also serves as an excellent toy model for 
understanding of the strongly coupled aspects of the real world 
QCD at finite temperature. The Debye correlation length
and the thermalization time scale of  finite 
temperature ${\cal N}=4$ super Yang-Mills theory in the strongly 
coupled regime are examples of such toy 
application\cite{Bak:2007fk,Bak:2007qw}. 

The type IIA counterpart is proposed recently, in which 
the ABJM theory
is three dimensional
 ${\cal N}=6$ U(N)$\times\overline{\mbox{U}}\mbox{(N)}$ superconformal 
Chern-Simons theory with level $(k, -k)$ and proposed to be dual to 
the type IIA superstring theory on AdS$_4\times \mathbb{CP}_3$ 
background\cite{ABJM}. Some test of this new duality has been carried out
based on the integrabilities with indication of some additional
structure\cite{Minahan:2008hf,Bak:2009mq}.
It is based on the large $N$ limit where one is taking 
$N,\  k\rightarrow \infty$  while holding the 't Hooft coupling
$\lambda={N\over k}$ to be fixed. If % the 't Hooft coupling 
$\lambda$ is small, 
the superconformal Chern-Simons theory %is weakly coupled and 
can be studied by perturbative analysis 
while, for strongly coupled regime of large $\lambda$, the supergravity 
is a valid description.

In this note, using the supergravity description, we shall study 
 thermal aspects of the ABJM field theory in the strongly coupled
regime. In order to identify the low energy dynamics, we first
carry out the consistent, $\mathbb{CP}_3$ invariant 
compactification of the type IIA supergravity down to  four 
dimensions. (See also Ref.~\cite{Gauntlett:2009zw} for a related M theory
dimensional reduction, in which  extra modes are present compared to
 its type IIA counterpart.)
 The resulting potential is minimized at the AdS$_4$ 
vacuum and depends on  three bulk scalars which are dual to
operators of dimension $4$, $5$ and $6$.
Using this 4d action, one may consistently study full nonlinear
effects without being limited to just small fluctuations. 

In addition the theory involves two bulk U(1) gauge fields satisfying Maxwell
type equations.  One combination is
the usual  massless U(1)  gauge  field which implies  existence
of global U(1) current of dimension $2$.
In the field theory side,  one has so called di-baryon charge\cite{ABJM} 
which
is %conserved and 
always accompanied by magnetic flux due to Gauss law 
constraint of Chern-Simons theories. Turning on the corresponding 
global charge will lead to a charged AdS black hole which is necessarily
deformed due to the nontrivial dependence on the scalars.
The other combination of bulk gauge fields turns out to be massive due 
to the Higgs mechanism where one scalar that is 4d hodge-dual to 
the NS-NS three form  gets absorbed into gauge degrees.
This massive gauge field is dual to the dimension $5$ current operator
in the boundary field theory side.

In this note, to understand 
the low energy dynamics of the ABJM system, 
we study the response of these operators
at zero and finite temperatures considering AdS$_4$ and 
the black brane backgrounds  respectively.

We shall demonstrate that, at zero temperature,
none of these operators develops  
a non-zero expectation value
under a constant source term. We also compute
exact expressions for AC conductivity of the two currents 
at zero temperature. At finite temperature phase, the dimension 
4 and 5 scalar operators possess  
nonvanishing expectation values meaning 
condensations. The magnitudes of condensation are proportional to
 source term and we shall compute these proportionality 
constants. 

The vector dynamics is particularly interesting. The source term 
coupled to boundary current operators is interpreted as a 
boundary vector potential. If the presence of this boundary vector potential 
implies  nonvanishing boundary current, this is precisely
the superconducting current which is proportional to the vector 
potential\cite{Hartnoll:2008vx,Basu:2008st}. 
The system is then in a superconducting phase.
For finite temperatures, we first show that
the supercurrent exists for the current of general 
dimensions with exceptions of $\Delta=3\,n\pm1\ (n\in \mathbb{N})$.
The $\Delta=2,\ 5$ cases belong to this %one of these
 exceptional category
and, hence, there exists no superconducting current for our black 
brane background.

The paper is organized as follows. Section 2 is a brief description of 
the type IIA supergravity and the AdS$_4\times\mathbb{CP}_3$
backgrounds. In Section 3, we carry out the $\mathbb{CP}_3$ invariant
dimensional reduction down to four dimensions. 
We pay a particular attention
to the consistency of ansatz with the original supergravity 
equations of motion. For completeness, we also include the dimensional 
reduction for the skew-whiffed background\cite{Duff:1984hn} corresponding to
the anti-D2 branes instead of the D2 branes of the ABJM background.
Section 4 describes the scalar dynamics and finite temperature condensation 
of dual field theory operators.
Section 5 describes the vector dynamics. We find that both dimension $2$
and $5$ currents are not superconducting currents. Last section
is devoted to concluding remarks and further directions of study.

%\nopagebreak
%%%%%%%%%%%%%%%%%%%%%%%%%%%%%%%%%%%%%%%%%%%%%%%%%%%%%%%%%%%%%%%%%%%%%
%%%%%%%%%%%%%%%%%%%%%%%%%%%%%

\section{Type IIA supergravity and the AdS$_4\times \mathbb{CP}_3$ 
background}

We begin our discussion from the type  IIA supergravity description 
of the ABJM field theory.
%The type IIA string theory on the AdS$_4\times \mathbb{CP}_3$
% background is proposed as a holographic dual of the 
%$N=6$  $\mbox{U(N)}\times \overline{\mbox{U}}
%\mbox{(N)}$ superconformal Chern-Simons
%theory with  %Chern-Simons 
%level $(k,-k)$. 
The supergravity description 
is valid in the limit $N\rightarrow \infty$ with
the 't Hooft coupling $\lambda={N\over k}$ kept fixed and large,  i.e.
$\lambda\ \gg\  1$.
%\,, \ \ \ \ \  \  {N\over k^5}\  \ll\  1\ee
%where $\lambda$ is the 't Hooft coupling given by $N/k$. 
Hence 
corresponding dual CFT is necessarily strongly coupled. Of course 
the small $\lambda$ region can be studied %described 
by  
direct  perturbative
analysis of the ABJM field theory.
The ABJM theory possesses $SO(3,2)\times SU(4)$ global bosonic 
symmetries.
%The bosonic global symmetries in the spacetime direction
The 4d conformal group of $SO(3,2)$ corresponds to 
%i, which is also the symmetries
%of the gravity background 
the global symmetry of AdS$_4$ spacetime. 
The Chern-Simons theory
also possesses  $SU(4)$ R-symmetry, which is the symmetry of
the $\mathbb{CP}_3$ directions. Their supersymmetric completion  is
described by the supergroup of $Osp(4|6)$. 
%Due to the presence of the $\mathbb{CP}_3$
The Chern-Simons theory has one additional global U(1) symmetry,
which we shall discuss later on.
We also consider  finite 
temperature version of this field theory 
in the large $\lambda$ regime, which can be 
described by the IIA supergravity in the AdS$_4$ black brane background.

In this section, we shall  briefly review the type IIA 
supergravity and some relevant thermodynamic properties of 
the AdS$_4$ black brane background.  
The bosonic part of the Einstein frame type IIA supergravity
is described by
the action,
\be
S_{IIA}= S_{NS}+ S_R+ S_{CS}
\ee
where
\bea
S_{NS}&=& {1\over 2 \kappa^2_{10}}\int
d^{10} x \sqrt{-g}\,\,\Big[\,\,
R(g)-{1\over 2}\nabla_m\phi\nabla^m\phi -{1\over 12}
e^{-\phi} \, H_{mnp}H^{mnp}\,\,
\Big] \nonumber\\
S_{R}&=&
{1\over 2 \kappa^2_{10}}\int
d^{10} x \sqrt{-g}\,\,\Big[
-{1\over 4}
e^{{3\over 2}\phi}\, F_{mn}\,\,F^{mn}
-{1\over 48}
e^{{1\over 2}\phi}\, \widetilde{F}_{mnpq} \,\widetilde{F}^{mnpq}
\Big]\nonumber\\
S_{CS}&=&
{1\over 2 \kappa^2_{10}}\int
%d^{10} x \sqrt{-g}
\,\,\Big[\,
\,
{1\over 2} \,\,
B_{[2]}\,\wedge\, F_{[4]}\,\wedge\, F_{[4]} \, \Big]
\,.
\eea
The 10d gravity coupling in this action is given by
$2k^2_{10}= (2\pi)^7 \, \ell_s^8 g_s^2$ where $g_s$ denotes
the string coupling.
We define the dilaton field $\phi$ as its nonzero mode by
subtracting its constant
part $\hat\phi$ related to the string coupling $g_s=e^{\hat{\phi}}$. 
%denotes the string coupling
%that is given by $e^{\hat{\phi}}$.
The gauge invariant 
four form field strength $\widetilde{F}_{[4]}$ is defined by
\be
\widetilde{F}_{[4]}= d A_{[3]} + dB_{[2]}\wedge A_{[1]}\,,
\ee
and the NS-NS three form field strength by $H_{[3]}=dB_{[2]}$.
The string frame metric is given by the transformation
$g^{s}_{mn}= e^{{1\over 2}\phi} g_{mn}$, but we shall not 
use it in this note.

The supergravity equations read
\bea
R_{mn}&=& {1\over 2} \nabla_m \phi\nabla_n \phi+
{1\over 2} %g_s^2 
\,e^{3\phi\over 2} \Big[F_{mp}F_{n}\,^p-{1\over 16} g_{mn}
\, F_{pq}F^{pq}\Big]\nonumber\\
&+&
{1\over 4}e^{-\phi}\Big[H_{mpq}H_{n}\,^{pq}-{1\over 12} g_{mn}
\, H_{pqr}H^{pqr}\Big]\nonumber\\
&+&{1\over 12} %g_s^2
\,e^{\phi\over 2}\,\Big[
\widetilde{F}_{mpqr}\widetilde{F}_{n}\,^{pqr}-{3\over 32} g_{mn}
\, \widetilde{F}_{pqrs}\widetilde{F}^{pqrs}
\Big]\,,
\label{eq24}
\eea
\vspace{-0.5cm}
\bea
&& 96\,\Box\,\phi =
36\, %g_s^2
\, e^{{3\over 2}\phi}\, F_{mn}F^{mn}
-{8}
e^{-\phi}\, H_{mnp}H^{mnp}
+
%g_s^2
\,e^{{1\over 2}\phi}\, \widetilde{F}_{mnpq}\widetilde{F}^{mnpq}\,,
\label{eqphi}\\
&&\nabla_m \big[
e^{3\phi\over 2} F^{mn}
\big]=-{1\over 6}e^{\phi\over 2} \widetilde{F}^{npqr}{H}_{pqr}\,,\\
&&\nabla_m \big[
e^{-\phi} H^{mnp}\big] +%g_s^2
{1\over 2}\,e^{\phi\over 2} \widetilde{F}^{mnpq}F_{mq}
=-{ 1 %g_s^2
\over 1152}\epsilon^{npr_1\cdots r_8}
 \widetilde{F}_{r_1\cdots r_4}\widetilde{F}_{r_5\cdots r_8}\,,
\label{eqh}
\\
&&\nabla_m \big[
e^{\phi\over 2} \widetilde{F}^{mnpq}\big]
-{1\over 144}\epsilon^{mnpq r_1\cdots r_6}
 \widetilde{F}_{r_1\cdots r_4} H_{mr_5 r_6}=0\,,
\label{Ein}
\eea
which agree with those in Ref.~\cite{Nilsson:1984bj}. 
%with an appropriate specification of constants.

The supergravity background
of the ABJM theory is given by\cite{Nilsson:1984bj,ABJM}
\be
\rmd s^2 &=&R^2_s  \left[ {1 \over 4} \rmd s^2
(\mbox{AdS}_4) + \rmd s^2 (\mathbb{CP}_3)  \right] \nonumber \\
e^{2 \hat\phi} &=& g_s^2=%{1\over k^2 }\,\,
{R_s^2 \over k^2\,\ell_s^2} 
\nonumber \\
F_{[4]} &=& {3 \over 8} k\, g_s R_s^2 \widehat{\epsilon}_4 \nonumber \\
F_{[2]} &=& k\, g_s\,\rmd \omega = 2 k \, g_s\, J \,,
\label{ABJbackground}
\ee
where the IIA curvature radius $R_s$
is
\be
R^2_s = %2^{5\over 2}
4 \pi \sqrt{2\lambda}\,\ell^2_s \,.
\ee
The unit-radius
AdS Poincar\'e metric is given by
\be
ds^2(\mbox{AdS}_4)= {1\over z^2}\,\,\Big[-dt^2 + dx^2 + dy^2 + dz^2\Big]\,,
\ee
and $\widehat{\epsilon}_4$ denotes its 4d volume form. 
%where $J$ is the K\"ahler two-form threading the $\mathbb{CP}_1$
%inside $\mathbb{CP}_3$. 
%where one can set the string length scale $\ell_s$ to be unity.
%Note that the 't Hooft coupling $\lambda=N/k$ is not a
%continuous parameter but
%a fractional number.
We parametrize the $\mathbb{CP}_3$ metric as\cite{Cvetic:2000yp}
\bea
\rmd s^2(\mathbb{CP}_3) &=& \rmd \xi^2 + {\sin^2 2\xi \over 4}
\left(
\rmd \psi + {\cos\theta_1\over 2} \rmd \phi_1 -
{\cos\theta_2\over 2} \rmd \phi_2
\right)^2
+ {1\over 4} \cos^2\xi
(\rmd \theta_1^2 +\sin^2\theta_1 \rmd \phi_1^2)\nonumber\\
&+& {1\over 4}\sin^2\xi (\rmd \theta_2^2 +\sin^2\theta_2 \rmd \phi_2^2) \, ,
\label{gravityabjm} 
\eea
where $0\le \phi_i < 2\pi$,  $0\le \theta_i \le \pi$,
$0\le \xi \le {\pi\over 2}$ and $0\le \psi < 2\pi$. 
%We shall not present the detailed parametrization of the
% $CP_3$ space but 
Note that the volume 
of unit  $\mathbb{CP}_3$ space is given by $\pi^3/6$.
The unit $\mathbb{S}^7$ can be presented as a Hopf
fibration over $\mathbb{CP}_3$,
\be
ds^2(\mathbb{S}^7) =\rmd s^2(\mathbb{CP}_3) +(d\theta_{10}+\omega)^2\,,
\ee
with $0\le \theta_{10} < 2\pi$.
The one form $\omega$ is explicitly given by 
\bea
\omega= - {1\over 2 %\sqrt{\lambda}
} \Big( {\cos 2\xi } \rmd \psi + \cos^2\xi
{\cos\theta_1} \rmd \phi_1 +\sin^2\xi
{\cos\theta_2 } \rmd \phi_2\Big)\,\,.
\eea
Then the K\"ahler two form $J_{\rm }={1\over 2}d\omega$
takes the form
\bea
%&& 
&& J = {1\over 4 %\sqrt{\lambda}
} \Big( \,\, {\sin 2\xi } \rmd \xi \wedge
\left(
 2\rmd \psi + {\cos\theta_1} \rmd \phi_1 -
{\cos\theta_2 } \rmd \phi_2
\right)%\Bigr.
%\nonumber\\
%&& \ \ \ \  \Bigr.
\nonumber\\
&& \ \ \ \ 
+ \cos^2\xi \sin\theta_1 \rmd \theta_1 \wedge \rmd \phi_1
+\sin^2\xi \sin\theta_2 \rmd \theta_2 \wedge \rmd \phi_2 \,\,
\Big)
 \,.
\eea
Below we shall use the following properties of $J$:
\bea
&&\nabla_a\, J_{bc}=0\,\,,\nonumber\\
&& J_{ab}\, J^{bc}=-\delta_{\,a}^{\,c}\,\,,\nonumber\\
&& 8 J^{ef}=\epsilon^{abcdef}\,J_{ab}\,J_{cd}\,\,,
\eea
where the indices are raised by unit $\mathbb{CP}_3$ metric.

We also consider the well known black brane solution 
with a planar symmetry. In this solution, 
one is replacing the AdS$_4$ metric in (\ref{gravityabjm}) 
by  the black hole metric
\be
ds^2_4= {1\over z^2}\,\,\Big[ {dz^2\over h(z)} - h(z)dt^2 + dx^2+dy^2\Big]
\ee
where
\be
h(z)= 1- \left({z\over z_H}\right)^3\,.
\ee
This black brane background is dual to the finite 
temperature version of the ABJM theory. Due to the quantum scale 
invariance of theory, this finite temperature field theory 
depends on only one dimensionful parameter which is the 
temperature $T$. Hence the theory possesses only one finite-temperature
 phase 
corresponding to  high temperature limit.
This temperature 
is identified with the Hawking temperature of the black brane,
\be
T={1\over 4\pi}|h'(z_H)|={3\over 4\pi}{1\over  z_H}\,.
\ee
The basic thermodynamic quantities are as follows; The entropy density
reads
\be
{\cal S}={N^2\over 3\sqrt{2\,\lambda}}\, {1\over z_H^2}
= {N^2\over 3\sqrt{2\,\lambda}}\, \Big({4\pi\, \over 3}\Big)^2\, T^2\,.
\ee
The energy density may be evaluated  as
\be
{\cal E}={N^2\over 6\pi \sqrt{2\,\lambda}}\, {1\over z_H^3}
= {2N^2\over 9 \sqrt{2\,\lambda}}\, \Big({4\pi\,\over 3}\Big)^2\, T^3\,,
\ee
and the pressure $p={\cal E}/2$ as dictated by the conformal symmetry.
We see here that the number of effective degrees in the 
strongly coupled regime %described by the gravity side 
is proportional to $N^2/\sqrt{\lambda}$. 
Since the number of degrees in the weakly coupled, 
small $\lambda$ region is simply proportional to $N^2$, the effective 
number of degrees is reduced greatly
down by the factor of $\sqrt{\lambda}$.

In order to simplify notation, we set up our convention as follows. 
%We normalize
%$g_s F_{[2]}\rightarrow F_{[2]}$ and
%$g_s F_{[4]}\rightarrow F_{[4]}$. 
We note that 
only $R_s$ dependence remains in the above 
background while
the ${k}$ dependence
disappears completely. The gravity 
equations themselves do not involve any explicit $k$ dependence either. 
Then we set $R_s=1$ taking $R_s$ as a length unit if necessary.

\section{Compactification on $\mathbb{CP}_3$}

We perform a consistent, $\mathbb{CP}_3$ invariant
dimensional reduction by taking the following 
ansatz\footnote{For the compactification spectrum, see
\cite{Duff:1984hn}-\cite{Halyo:1998pn}.
 See also Ref.~\cite{Gauntlett:2009zw} for 
 the related dimensional reduction from the M theory  whose
spectra are different 
% The part involving  gauge fields includes
% non-trivial transformations, which makes the  comparison
%non-straightforward. 
from those of  the type IIA theory.
}:
\bea
\rmd s^2 &=& %\left[ 
{1 \over 4} e^{-3\sigma}\rmd s^2_4 +
e^{\sigma} \rmd s^2 (\mathbb{CP}_3) % \right]
\,\,\,, \nonumber \\
F_{ab} \ \ &=& 2 J_{ab}\,\,,\ \ \ \ \ F_{\mu\nu}=\partial_\mu A_\nu 
-\partial_\nu A_\mu \,\,,
\nonumber \\
H_{\mu\nu\lambda} &=& {3\over 2}\, e^{{\phi}-6\sigma}\,
\epsilon_{\mu\nu\lambda}\,^{\delta}
% \sqrt{-g_4}\,\, E_{\mu\nu\lambda\delta}\,
(\pm A_\delta-\bar{A}_\delta-\nabla_\delta \psi)\,,\ \ \ \ \ \ \ 
H_{\mu ab}=(\partial_\mu \chi)\,\, J_{ab} \,\,\,,
\nonumber\\
\tilde{F}_{\mu\nu\lambda\delta} &=&- {3 \over 8}\, e^{-{\phi\over 2}-9\sigma}\,
\epsilon_{\mu\nu\lambda\delta}
%  \sqrt{-g_4}\,\, E_{\mu\nu\lambda\delta}
\,\,(\pm 1+\chi^2)\,\,,
\nonumber \\
\tilde{F}_{\mu\nu ab} &=& e^{-{\phi\over 2}-\sigma}
\,\,^*\tilde{F}_{\mu\nu}\,\, J_{ab}\,\,,
\ \ \ \ \tilde{F}_{abcd}= -2 \chi \, (J\wedge J)_{abcd}\,,
\label{cp3ansatz}
\eea
where we use  $\epsilon_{\mu\nu\lambda\delta}=
\sqrt{-g}\,E_{\mu\nu\lambda\delta}$ 
with a convention $E_{0123}=-1$ for the numerical totally 
antisymmetric
number $E_{\mu\nu\lambda\delta}$.
The 4d  hodge dual is then defined by  $\,^*F_{\mu\nu}={1\over 2}
\epsilon_{\mu\nu}\,^{\alpha\beta}\,\, F_{\alpha\beta}$ and 
$\tilde{F}_{\mu\nu}$ is simply an antisymmetric tensor field
not defined by the vector potential. 
We shall  turn off all the remaining components.
The upper signs in the above ansatz and below correspond to the
ABJM theory whereas the lower signs are for the skew-whiffed 
background\cite{Duff:1984hn} corresponding to the anti-D2 branes. 
From \cite{Nilsson:1984bj}, one can check that all of $\mathbb{CP}_3$
invariant modes are included in the above  ansatz. Note also that there are no
$\mathbb{CP}_3$ invariant fermionic modes\cite{Nilsson:1984bj}.

The resulting equations of motion can be derived from
the action
\be
{\cal L}_{4}= {1\over 16 \pi G_4}\,\,\big(
\,{\cal L}_g +{\cal L}_c\,
\big)\,\,,
\label{effaction}
\ee
where
\bea
&& {\cal L}_g= % {1\over 16 \pi G_4}
%\int
%d^{4} x \sqrt{-g}\,
%\,\Big[\,\,
R_4(g)-{1\over 2}(\nabla\phi)^2
-{6}(\nabla\sigma)^2-{3\over 2}e^{-\phi-2\sigma}\, (\nabla\chi)^2
-e^{{3\phi\over 2}+3\sigma} \, F_{\mu\nu}F^{\mu\nu}-3
e^{-{\phi\over 2}-\sigma} \, \tilde{F}_{\mu\nu}\tilde{F}^{\mu\nu}
 \nonumber\\
&&\ -18 e^{{\phi}-6\sigma}(\pm A-\bar{A}-\nabla\psi)^2+
12e^{-4\sigma}-{9\over 2} e^{-{\phi\over 2}-9\sigma}\,(\pm 1+\chi^2)^2
-{3\over 2} e^{{3\phi\over 2}-5\sigma}-6\,\chi^2 
e^{{\phi\over 2}-7\sigma}\!,
\eea
and
\bea
{\cal L}_c&=&6\,e^{-{\phi\over 2}-\sigma}\Lambda^{\mu\nu}
\Big(\,\,
\tilde{F}_{\mu\nu}-2\chi e^{-{\phi\over 2}-\sigma}\,\,\,^* 
\tilde{F}_{\mu\nu}-\chi^2 F_{\mu\nu}-\partial_\mu \bar{A}_\nu+
\partial_\nu \bar{A}_\mu\,\,
\Big)\nonumber\\
&+& 6\chi e^{-\phi-2\sigma}\,
 \tilde{F}_{\mu\nu}\,^*\tilde{F}^{\mu\nu}+6\,\chi \,
 \tilde{F}^{\mu\nu}\big(\,\,^*{F}_{\mu\nu}+2\,\chi\, e^{-{\phi\over 2}-\sigma}
\,F_{\mu\nu}
\,\big)-4 \chi^3 \,{F}_{\mu\nu}\,^*{F}^{\mu\nu}\,.
\eea
The $g_{\mu\nu}$, $\phi$ and $\sigma$ equations  are solely following 
from the variations of ${\cal L}_g$. Namely the corresponding 
variations of ${\cal L}_c$ are vanishing completely.
% as discussed in Appendix. 
The $\Lambda^{\mu\nu}$ variation leads to the constraint
\be
\tilde{F}_{\mu\nu}-2\chi e^{-{\phi\over 2}-\sigma}\,\,\,^* 
\tilde{F}_{\mu\nu}-\chi^2 F_{\mu\nu}- \bar{F}_{\mu\nu}=0\,.
\label{cons}
\ee
The $\tilde{F}_{\mu\nu}$ variation determines $\Lambda_{\mu\nu}$ by
\be
\Lambda_{\mu\nu}=\tilde{F}_{\mu\nu}-\chi \,  e^{{\phi\over 2}+\sigma}
\,\,^* F_{\mu\nu}\,,
\ee
and the remaining equations are 
\bea
\nabla_\mu \big[\,\, e^{{3\phi\over 2}+3\sigma}\, F^{\mu\nu}\,\,\big]
&=& 3(\partial_\mu \chi)\,\,^* \tilde{F}^{\mu\nu}
+9\, e^{{\phi}-6\sigma}\,(\pm 1+\chi^2)\, 
\big(\,\pm A^\nu-\bar{A}^\nu-\nabla^\nu\psi\,\big)\,,\nonumber\\
\nabla_\mu \big[\,\, e^{-{\phi\over 2}-\sigma}\, \tilde{F}^{\mu\nu}\,\,\big]
&=& (\partial_\mu \chi)\,\,^* {F}^{\mu\nu}
-3\, e^{{\phi}-6\sigma}\,
\big(\,\pm A^\nu-\bar{A}^\nu-\nabla^\nu\psi\,\big)\,,\nonumber\\
\nabla_\mu \big[ e^{-{\phi}-2\sigma}\, \nabla^\mu\,\chi\big]
&=& 
4\,\chi\,
e^{{\phi\over 2}-7\sigma} +
6\chi\,(\pm 1+\chi^2)\, e^{-{\phi\over 2}-9\sigma}-
2 \,\,^* \tilde{F}^{\mu\nu}\,\big(\,
{F}_{\mu\nu} -e^{-\phi-2\sigma}\tilde{F}_{\mu\nu}
\,\big)\,.
\eea
Details of derivation  are relegated to Appendix.
The constraint (\ref{cons}) can be solved in terms of vector  
potentials by
\be
\tilde{F} =
{1\over 1+4\chi^2 e^{-\phi-2\sigma}}\Big[
\,\,\bar{F}+\chi^2 F +2\chi e^{-{\phi\over 2}-\sigma}\,\,
\,^* \big(\bar{F}+\chi^2 F\big)
\,\,\Big]\,.
\ee
For the field theory interpretation, one needs the 4d Newton constant
whose value is given by
\be
{1\over 16 \pi G_4}= {N^2\over 12 \pi \sqrt{2\lambda}}\,.
\ee
%But in the end, we shall translate the mass unit in terms
%of $T$ by multiplying $4\pi/3$.

As identified in Ref.~\cite{ABJM}, 
the field $F_{\mu\nu}$ 
electrically couples to D0 brane current  
while its magnetic hodge dual couples to  D6
branes wrapping $\mathbb{CP}_3$.  
The other field $\tilde{F}_{\mu\nu}$
couples electrically to  D4 branes wrapping $\mathbb{CP}_2$ cycle 
and magnetically to D2 wrapping $\mathbb{CP}_1$ cycle of $\mathbb{CP}_3$.  

%The Euclidean black hole background is given by
%\be
%ds^2_4= {1\over z^2}\Big(h(z)d\tau^2+{dz^2\over h(z)}+dx^2+dy^2\Big)\,,
%\ee
Finally, we note that all the 4d equations are $R_s$ independent and scale
invariant. One may further set the only length scale $z_H=1$ 
in the black brane metric 
as $h(z)=1-z^3$. This corresponds to the mass unit
${4\pi T}/3$, which will be recovered whenever needed.
Below we shall be only interested in the case of the  ABJM background 
for the further analysis.

\section{Scalar dynamics}

There are three scalars, $\phi$, $\sigma$ and $\chi$ 
for our low energy effective action. Unlike the AdS$_5$ case of the 
type IIB 
theory, all of them become massive. In this section we shall explore
their zero and finite temperature dynamics. 
%From their behaviors near the boundary 
%at $z=0$, one may turn on corresponding operators   

Under $\phi$ and $\chi$ changes, the gravity potential in 
(\ref{effaction}) becomes infinitely large when 
$|\phi|$ and $|\chi|\ \rightarrow\ \infty$. The potential has an 
absolute minimum at $\phi=\sigma=\chi=0$, at which the potential 
 takes a %has the minimum 
value $-6$.
When $\sigma\rightarrow -\infty$, the potential  becomes 
infinitely large as well. In the other limit $\sigma\rightarrow \infty$,
%$V\rightarrow 0$.
the potential goes to zero.
 The latter corresponds to the 
decompactification limit where the volume of $\mathbb{CP}_3$ becomes
infinitely large.

For the small fluctuations to the linear order, 
the $\phi$ and $\sigma$ fields can be diagonalized
by the linear combinations,
\be
\phi_+={%\big(
\phi+18\sigma %\big)
\over
\sqrt{28}}\,,\ \ \ \  \phi_-={%\big(
 \sqrt{3}\big(3\phi -2%\sqrt{3}
\sigma\big)
\over
\sqrt{28}}\,,
\ee
which lead to the massive scalar equation,
\be
\nabla^2 \Phi - m_\Phi^2\, \Phi=0\,,
\ee
with masses $m_{\phi_+}^2=18$ and $m_{\phi_-}^2=4$ 
respectively. 
The $\chi$ field, on the other hand, is already diagonal
with mass $m^2_\chi=10$. As is well known,
scalar field in the AdS$_4$ space behaves, near the boundary 
$z=0$, as\cite{Balasubramanian:1998de}
\be
\Phi(x,z)
 = a_\Phi(x) \, \, z^{3-\Delta}+ b_\Phi(x)\,\,  z^{\Delta}+\cdots\,,
\label{bdb}
\ee
where $\cdots$ denotes all the remaining powers of $z$
and $x_{\bar{\mu}}$ is for the boundary directions $x_0,x_1,x_2$.
The number $\Delta$ in the above expression is given by
\be
\Delta={1\over 2}\Big(\,{3+\sqrt{9+4m^2_\Phi}}\,\,
\Big)\,,
\ee
which can be identified with 
the scaling dimension of the field theory operator 
dual to $\Phi$. It follows that the dimensions of
the dual operators $O_{\phi_+}$, $O_{\phi_-}$ and
$O_{\chi}$ are respectively $6$, $4$ and $5$.
This is quite consistent with the fact that 
these operators should belong to some low-lying protected
supermultiplets where  dimensions of
component operators do not receive any quantum 
corrections such that their bosonic sector consisting of even number 
of fields have 
only integer dimension.

Turning on $a_\Phi$ in (\ref{bdb}) in the supergravity side
corresponds to turning on the source term
in the field theory side,
\be
\delta {\cal L}_{FT}= a_\Phi(x)\,\, O_\phi(x)\,.
\ee
Then the operator expectation value $\langle O_\Phi(x) \rangle$
is evaluated as
\be
\langle\, O_\Phi(x) \,\rangle={\delta\, I_{SUGRA}\over {\delta a_\Phi}}
={3\,\alpha\over 32\pi G_4}\,\, b_{\Phi}(x)\,,
\ee
where $\alpha$ is an extra factor in the scalar kinetic term
$-{(\nabla\Phi)^2/2}$ in ${\cal L}_g$ related to the 
normalizations of our scalar field.
The boundary condition for the scalar field
at the horizons at $z=1$ or $z=\infty$ (\,$T=0$\,) is
\be
%\sqrt{-g}\,
 h\, %g^{zz}
\, \Phi'\,|_{z=1\,,\,\, \infty}=0\,,
\ee
where  prime denotes a derivative with respect to $z$.
This is the requirement that there should not be any boundary 
contribution from the horizon when one evaluates the on-shell 
supergravity action\footnote{In principle, $\Phi=0$ boundary 
condition is allowed but this in general leads to null 
solution besides some special cases which are 
not relevant to us.}. 

Now we consider the massive scalar equation with an 
ansatz $\Phi= U(z)$ in order to study the finite temperature 
condensation of operators.
For $T=0$, the scalar equation becomes
\be
z^2\, U'' -2 z\, U' - m^2_\Phi\, U=0\,,
\ee
which leads to the simple solution,
\be
U= a_\Phi \, \, z^{3-\Delta}+ b_\Phi\,\,  z^{\Delta}\,.
\ee
For $\Delta=4,5,6$ of our problem,
the boundary condition at %the horizon 
$z=\infty$ demands
$b_\Phi=0$. Hence there is no zero-temperature  
condensation of operators 
%when we turns on source terms
by coupling of source term
in the field theory side.  

The equation for $T\neq 0$ becomes
\be
z^4\,\Big({h\over z^2}\, U'\,\Big)'- m^2_\Phi U=0\,.
\ee
This leads to the general solution
\be
U=a_\Phi \, \, z^{3-\Delta}\,\,_2 F_1
\Big(\,{3-\Delta\over 3},\,
{3-\Delta\over 3};\,
{6-2\Delta\over 3}
;\,
z^3\Big)+ b_\Phi\,\,  z^{\Delta}
\,_2 F_1
\Big(\,{\Delta\over 3},\,
{\Delta\over 3};\,
{2\Delta\over 3}
;\,
z^3\Big)\,,
\ee
where $\,_2 F_1(\alpha,\beta;\gamma;\,x)$ is the hypergeometric function. 
For $\Delta =6$ of $m^2_{\phi_+}=18$, the hypergeometric function can be 
simplified to
\be
U=a_{\phi_+} \, \Big({1\over z^3}-{1\over 2}\Big)
-6\, b_{\phi_+}\, \Big(
2 +(2 z^{-3} -1)\ln(1-z^3)
\Big)\,.
\ee
The boundary condition at the horizon  dictates then
\be
{b_\Phi\over a_\Phi}=
-{4\Gamma\Big({3\over 2}-{\Delta\over 3}\Big)
\, \Gamma\Big({\Delta\over 3}\Big)\over 
4^{2\Delta\over 3}\,\Gamma\Big({3-\Delta\over 3}\Big)
\, \Gamma\Big({1\over 2}+{\Delta\over 3}\Big)
}\,\Big({4\pi\over 3}T\Big)^{2\Delta-3}\,,
\ee
where our mass unit is recovered.
 For $\Delta={3\over 2}+3n$ 
($n=1,2,\cdots$),
one has $a_\Phi=b_\Phi=0$, which means 
turning on the operator is not allowed.
For the other case $\Delta=3n$ ($n=1,2, \cdots$), one finds
$b_\Phi=0$. Hence there is no condensation of the corresponding
operators.
%in the presence of  source. 
The $\phi_+$ fluctuation 
belongs to this category where $\Delta=6$.
The massless case of  $\Delta=3$ is special; Both 
$a_\Phi$ and $b_\Phi$ are allowed and independent of each other.
For the generic  mass, the above ratio is 
nonvanishing. In particular, the  $\phi_-$ and 
$\chi$ fluctuations belong to this category leading to 
 finite-temperature condensations. 

For $m^2_{\phi_-}=4$ with $\Delta=4$, one has
\bea
\langle\,O_{\phi_-}\,\rangle
&=&{3\over 32\pi G_4}\,\,{\sqrt{3}\,
\big[\,\Gamma\big({1\over 3}\big)\,\big]^6
\over 160 \pi^3}\, \Big({4\pi\over 3}\,T\Big)^5 \,\, a_{\phi_-}
= {2\,\pi\,\sqrt{6}\,\big[\,\Gamma\big({1\over 3}\big)\,\big]^6
\over 1215 }\, {N^2 T^5\over {\sqrt{\lambda}}}
\,\, a_{\phi_-}
\nonumber\\
&=&(4.6823)\,  
{N^2 T^5\over {\sqrt{\lambda}}}\,\, a_{\phi_-}\,,
\eea
where we recover our mass unit.
For $m^2_{\chi}=10$ with $\Delta=5$, one has 
\bea
\langle\,O_{\chi}\,\rangle
&=&-{9\over 32\pi G_4}\,\,{9\sqrt{3}
\, \big[\,\Gamma\big({2\over 3}\big)\,\big]^6
\over 56 \pi^3}\, \Big({4\pi\over 3}\,T\Big)^7 \,\, a_{\chi}
=- {128\,\sqrt{6}\,\pi^3\,\big[\,\Gamma\big({2\over 3}\big)\,\big]^6
\over 567}\, {N^2 T^7\over {\sqrt{\lambda}}}\,\, a_{\chi}\nonumber\\
&=&-(105.70)\,  
{N^2 T^7\over {\sqrt{\lambda}}}\,\, a_{\chi}\,.
\eea
Therefore at finite temperature, the condensations of
$O_\chi$ and $O_{\phi_-}$ occur
if one dials the corresponding source terms. 
%in the ABJM field theory system.  
This also verifies that
the $T\neq 0$ phase of the ABJM  theory is 
distinct from that of $T=0$.
%the zero temperature phase.

\section{Vector dynamics and superconducting current}
We now turn to the analysis of the vector fields, $A_\mu$ 
and $\bar{A}_\mu$. At the linear order, $\tilde{F}_{\mu\nu}
\sim \bar{F}_{\mu\nu}$ and the combination, $A_\mu-\bar{A}_\mu$\,, 
becomes massive due to the Higgs mechanism. The absorbed
scalar $\psi$ is four dimensional 
 hodge dual to the three-form field 
$H_{\mu\nu\lambda}$. 
Including this combination, 
the gauge kinetic 
terms can be diagonalized by the linear combinations,
\be
A^H_\mu={\sqrt{3}\over 2}\big(A_\mu-\bar{A}_\mu\big)\,,
\ \ \ \ \ A^B_\mu={1\over 2}\big(\, A_\mu+3\,\bar{A}_\mu\big)\,,
\ee
with masses $m^2_H=12$ and $m^2_B=0$.
The massless field $A_\mu^B$ couples to
the di-baryon current $J^B$ of the ABJM theory. The field equation
including Gauss law constraint takes a form,
\be
{k\over 4\pi} {\ast}_3 {\rm tr}\big(\,F_{\mbox{\scriptsize $\rm U(N)$}}+
F_{\mbox{\scriptsize $\rm \overline{U}(N)$}}\,\big)= J_B
\ee
where ${\ast}_3$ denotes three dimensional hodge dual of the boundary 
theory and $F_{\rm U(N)}$ and $F_{{\rm \overline{U}(N)}}$
are the 3d field strengths of U(N) and $\rm \overline{U}(N)$ gauge 
fields. Hence the charge of $J_B$ is always accompanied
by magnetic flux, which is an important characteristic of 
Chern-Simons theories.
Since the dual AdS gauge field $A^B_\mu$ is massless,
one may turn on the di-baryon charge and consider 
 charged AdS black hole background. %The presence of 
The scalar fields, 
$\phi,\sigma$ and $\chi$, can be consistently set to zero leading to 
a simple charged 
(dyonic) black hole solution with 
$A_\mu=\bar{A}_\mu$\cite{Gauntlett:2009zw}. 
%should be necessarily deformed.
%At least at linear order, existence of such black hole 
%solution is obvious. 
We shall get back to this issue 
 later on.  

To the linear order the gauge field equation satisfies
\be
\nabla_\mu\, F^{\mu\nu}\,- m^2_\varphi\, A^\nu=0\,.
\ee
If $m^2_\varphi\neq 0$, the consistency of the above equation
requires $\nabla_\mu A^\mu=0$, which also follows from the 
scalar equation $\psi$ in the gauge $\psi=0$. The massless 
case is special but the treatment is basically the same.

As in the scalar case, the fields, in the near boundary region
of $z=0$, behave as  
\be
A_{\bar\mu}= a_{\bar\mu} (x)\, z^{2-\Delta} + 
b_{\bar\mu} (x) \, z^{\Delta-1}+\cdots
\ee
where barred indices are  for the boundary spacetime directions,
i.e. $\bar{\mu}=0,1,2$. The number $\Delta$, % that is given by
\be
\Delta ={1\over 2}\,\Big(\,3+\sqrt{1+4\,m^2_\varphi}\,\Big)\,,
\ee
is the dimension of current operator dual
to  bulk gauge field $A_\mu$. For our cases of $m^2_\varphi=0$ and $12$,
we find $\Delta=2$ and $5$ respectively. The dimension two 
current operator
is the di-baryon current $J_B$ as we already mentioned. The other
dimension five current is dual to the massive gauge field, which
includes dynamics of the absorbed
scalar degree.  

We begin with the zero temperature case. With an ansatz
$A_{\bar\mu}={\cal A}_{\bar\mu}(z)$, the Maxwell equation becomes
\be
z^2\,{\cal A}''_{\bar{\mu}} -m^2_\varphi \, 
{\cal A}_{\bar{\mu}}=0\,,
\ee
whose solution is simply
\be
{\cal A}_{\bar{\mu}}= a_{\bar\mu} \, z^{2-\Delta} + 
b_{\bar\mu}  \, z^{\Delta-1}\,.
\ee
The vanishing boundary contribution at $z=\infty$ leads to
$b_{\bar\mu}=0$ for $\Delta\ge 2$. Consequently,
there is no induced
current by couping of constant vector potential
$a_{\bar\mu}$.

We now turn to  AC  response of current at zero temperature.
For this purpose, we shall consider only transverse current with
an ansatz, ${A}_y={\cal A}_y(z)\, e^{-i\omega\, t+ipx}$.
The Maxwell equation is reduced to %becomes
\be
z^2\, {\cal A}''_y +\Big(\,
(\omega^2-p^2)\,z^2 -m^2_\varphi\,\Big){\cal A}_y=0\,,
\ee 
whose general solution is
\be
{\cal A}_y =C_1\,z^{1\over 2}\, I_{\nu}\, \Big(\,z\,\sqrt{p^2-\omega^2}\,\Big)
+C_2\,z^{1\over 2}\, K_{\nu}\,\Big( \,z\,\sqrt{p^2-\omega^2}\,\Big)\,,
\ee
with $\nu=\Delta-{3\over 2}$. For $p^2 > \omega^2$, $I_\nu(x)$ becomes
exponentially large at the horizon $z=\infty$ so that
one requires $C_1=0$. If $p^2 < \omega^2$, one may 
use the so-called purely ingoing boundary condition
where flux near horizon region should be directed toward
the horizon. This again leads to $C_1=0$. Since $K_\nu(x)$
behaves
\be
K_\nu(z)= {\pi\over 2\, \sin\,(\nu\pi)}
\Big[\, {z^{-\nu}\over 2^{-\nu}\, \Gamma(1-\nu)}-
{z^\nu\over 2^\nu\, \Gamma(1+\nu)}
\Big]+\cdots
\ee 
for the small $z$, we conclude
\be
{b_y\over a_y}={\pi \,\,(\omega^2-p^2)^{\Delta-{3 \over 2}}
\over 
(2i)^{\,2\Delta-3}\,\, \Gamma\Big(\Delta-{1\over 2}\Big)
\Gamma\Big(\Delta-{3\over 2}\Big)\, \sin\big(\Delta-{1\over 2}\big)\pi
}\,.
\ee
For the massless case of $\Delta=2$, the AC conductivity
for the baryon current is then evaluated as
\be
\sigma_B={J_y\over i\omega \, a_y}=
{1\over 8\pi G_4}\, {b_y\over i\omega a_y}
={N^2\over 6\pi\sqrt{2\lambda}}\, \Big(1-{p^2\over \omega^2}\Big)^{1\over 2}\,,
\ee
while
\be
\sigma_H=\,{N^2\over 9450\,\pi\,\sqrt{2\lambda}}\, \omega^6 
\Big(1-{p^2\over \omega^2}\Big)^{{7\over 2}}\,.
\ee
The overall dependence on the frequency is basically dictated by the 
underlying conformal symmetry and  dimension of the current 
operator.

For the finite temperature, we again consider turning on
vector potential that is independent of the 
boundary coordinates. The Lorentz symmetry is now broken
due to the non-zero temperature but the rotation and 
translation symmetries remain. We take an ansatz
$A_{\bar\mu}={\cal A}_{\bar\mu}(z)$ as in $T=0$ case. 
For ${\cal A}_i\,\,\, (i=1,2)$, the Maxwell
equation becomes
\be
z^2 \Big(\, h\, {\cal A}'_i\Big)' -m^2_\varphi\,\, {\cal A}_i=0\,.
\ee
While turning on constant vector potential, the existence 
of the current proportional to the vector potential
as $J_i\propto a_i$ corresponds to  superconducting 
current\cite{Basu:2008st}. 
%If this is the case, the phase will be a 
Existence of such current implies that the system under 
consideration is in a
superconducting phase. We shall focus on 
the investigation of this possibility within our dimensionally reduced
system\footnote{
For the discussion of 
 superconductivity arising in the more general dimensional 
reduction, see Ref.~\cite{Gauntlett:2009zw,Gauntlett:2009dn}.}. 

The general solution of the above equation can be given by
\be
{\cal A}_i= a_i \, \, z^{2-\Delta}\,\,_2 F_1
\Big(\,{2-\Delta\over 3},\,
{4-\Delta\over 3};\,
{6-2\Delta\over 3}
;\,
z^3\Big)+ b_i\,\,  z^{\Delta-1}
\,_2 F_1
\Big(\,{\Delta-1\over 3},\,
{\Delta+1\over 3};\,
{2\Delta\over 3}
;\,
z^3\Big)\,,
\ee
in terms of the hypergeometric functions.
The boundary condition $h\, A'_i=0$ at $z=1$ leads to the ratio,
\be
{b_i\over a_i}=
-{\Gamma\Big({6-2\Delta\over 3}\Big)
\, \Gamma\Big({\Delta-1\over 3}\Big)
\, \Gamma\Big({\Delta+1\over 3}\Big)
\over 
\Gamma\Big({2-\Delta\over 3}\Big)
\, \Gamma\Big({4-\Delta\over 3}\Big)
\, \Gamma\Big({2\Delta\over 3}\Big)
}\, \Big({4\pi\over 3}\, T\Big)^{2\Delta-3}\,.
\ee
This is in general nonvanishing with some exception specified below 
and leads to the nonvanishing supercurrent in the presence of 
constant vector potential in the boundary theory.
The exception corresponds to
$\Delta =3n-1,3n+1$ where $n$ is a natural number.
Interestingly, our cases of $\Delta=2,\  5$ belong to this exceptional 
category. Therefore there is no superconducting current implying 
that our system is not in a superconducting phase. Note also 
that the hypergeometric functions for $\Delta=2,\ 5$ are greatly simplified
and  can be given in terms of simple elementary functions. But we shall
not present their detailed forms here.

For the time component, the equation is reduced to
\be
z^2 \, h\, {\cal A}''_0 -m^2_\varphi\,\, {\cal A}_0=0\,.
\ee 
whose solution is 
\be
{\cal A}_0= a_0 \, \, z^{2-\Delta}\,\,_2 F_1
\Big(\,{2-\Delta\over 3},\,
{1-\Delta\over 3};\,
{6-2\Delta\over 3}
;\,
z^3\Big)+ b_0\,\,  z^{\Delta-1}
\,_2 F_1
\Big(\,{\Delta-1\over 3},\,
{\Delta-2\over 3};\,
{2\Delta\over 3}
;\,
z^3\Big)\,,
\ee
This is not a valid solution when 
$\Delta=3\,n\ (\,n\in \mathbb{N}\,)$. But we shall not consider
$\Delta=3\,n$ since they are not relevant to our 
analysis. The boundary 
condition required for the vanishing contribution at the horizon
can be identified as $A_0=0$ unless $\Delta=2$. This leads to 
the ratio
\be
{b_0\over a_0}=
-{\Gamma\Big({6-2\Delta\over 3}\Big)
\, \Gamma\Big({\Delta+1\over 3}\Big)
\, \Gamma\Big({\Delta+2\over 3}\Big)
\over 
\Gamma\Big({4-\Delta\over 3}\Big)
\, \Gamma\Big({5-\Delta\over 3}\Big)
\, \Gamma\Big({2\Delta\over 3}\Big)
}\, \Big({4\pi\over 3}\, T\Big)^{2\Delta -3}\,.
\ee
The source term $a_0$ can be related to a chemical potential
while $b_0$ is an induced charge density. Hence non-zero 
chemical potential induces corresponding charge density
%Then again there exist superconducting current component
unless $\Delta=3\, n+2, 3\,n+1\ (\,n\in \mathbb{N}\,)$.
%which agrees with the analysis of the spatial components.
For $\Delta=3\, n+2, 3\,n+1\ (\,n\in \mathbb{N}\,)$, the ratio 
vanishes and there is no induced charge density. 
$J^0_H$ belongs to this exceptional category.
Thus it is not possible
to turn on a nonvanishing charge density  
 by adjusting the chemical potential.
%$J^H$ is not a superconducting current.

Finally for $\Delta=2$, the solution simply becomes
\be
A_0 = a_0^B+ b_0^B\, z
\ee
where $a_0^B$ is a constant which can be shifted away and
$b_0^B$ is the di-baryon charge density which can be freely adjusted. 
Hence this fluctuation 
describes the deformation of charged AdS black hole at linear order.

\section{Conclusions}

In this note, we considered
the thermal aspects of the ABJM field theories in the strongly 
coupled regime using the supergravity description. For this 
purpose, we carry out the $\mathbb{CP}_3$ invariant dimensional 
reduction of type IIA  supergravity down to four dimensions.
We then study the zero and finite temperature
responses of various operators which are dual to bulk scalar and 
vector fields. We have shown that
condensations of dimension $4$ and $5$ scalar operators
occur at finite temperature by coupling of constant source
term. The currents, on the other hand, are not induced 
by coupling of boundary vector potential, which implies 
that the system is 
not in a superconducting phase. We have also computed the AC 
conductivities at zero temperature.
 
The existence of  condensation
implies that the finite temperature phase is distinct
from the zero temperature phase of the ABJM theory.
This is rather obvious since the temperature breaks the 
scale invariance introducing  dimensionful 
temperature scale. The situation may turn into more interesting 
case if one dials the global U(1) charge that is dual to the
massless bulk gauge field. It is described by the 
charged AdS black hole, which is known to have the zero temperature
limit.  Unlike the case of the R-charged AdS$_5$ black 
holes\cite{Chamblin:1999tk,Bak:2006dn},
the zero temperature phase appears to be thermodynamically 
stable. Its phase structure and thermodynamic
properties are of interest\cite{Fujita:2009kw}.
The relation between charge and energy near zero 
temperature
\cite{Rey:2008zz,
Bak:2009kz} and the peculiar ground state entropy demand
a further understanding of the ABJM system with finite
di-baryon charge density.
The scalar $\chi$ and the gauge field $A^H$, both are dual
to the dimension 5 operators, get mixed
up if one turns on (dyonic) charges and may lead to a very nontrivial 
response
properties. 
These issues are currently under investigation.

The other is on the spatial correlation length 
scales of the ABJM 
theory in the strongly coupled regime. As discussed in 
Ref.~\cite{Bak:2007fk}, the longest correlation is governed by the true mass 
gap and the Debye mass for the charged excitation  
is the lowest in the CT-odd sector of 
the spectrum.  
Our four dimensional  gravity
action is particularly suited for this problem. This is also currently 
under investigation\cite{kazem}.

\section*{Acknowledgement}
This work was supported in part by 
NRF
R01-2008-000-10656-0 and 
NRF SRC-CQUeST-2005-0049409.

\appendix

\section{$\mathbb{CP}_3$ invariant dimensional reduction}

In this appendix, we shall explain 
some key ingredients of the
%the basic points to perform 
dimensional reduction and show that %it is possible to derive all 
all the equations of motion may be derived from the four dimensional 
effective  Lagrangian. The starting point is to solve %consistently 
the Bianchi 
identity, $d\tilde{F}_{[4]}=-H_{[3]}\wedge F_{[2]}$. We take the ansatz 
where %consider only the 
%case where 
$\tilde{F}_{\mu\nu\lambda a}=\tilde{F}_{\mu a b c}=
H_{\mu\nu a}=0$ and $F_{a b}=2J_{a b}$. By setting 
$\tilde{F}_{a b c d}=-2\chi\,(J\wedge J)_{a b c d}$ together with 
$\tilde{F}_{\mu\nu a b}=e^{-{\phi\over 2}-\sigma}
\,\,\,^\ast\tilde{F}_{\mu\nu}\,J_{a b}$,  we get
\be
H_{\mu a b}=\nabla_{\mu}\chi\,J_{a b} \,\,\,\,\,\textrm{and}
\,\,\,\,\, \nabla_\mu (\,e^{-{\phi\over 2}-\sigma} \tilde{F}^{\mu\nu}\,)=
\nabla_\mu\chi\,\,\,^\ast F^{\mu\nu}-{1\over 3}\epsilon^{\nu\mu\lambda\rho}
\,H_{\mu\lambda\rho}\,,
\ee
from the Bianchi identity.
With a further ansatz, 
$\tilde{F}_{\mu\nu\lambda\rho}=X e^{-{\phi\over 2}-9\sigma} 
\,\epsilon_{\mu\nu\lambda\rho}$, the Bianchi identity %eq.(\ref{Fmnpq})
% for %$\tilde{F}^{\mu\nu\lambda\rho}$ 
%$\mu\nu\lambda\rho$ directions
can be solved by
\be
\tilde{F}_{\mu\nu\lambda\rho}=-
{3\over 8}(\pm 1+\chi^2)\, e^{-{\phi\over 2}-9\sigma} 
\,\epsilon_{\mu\nu\lambda\rho}\,.
\ee
%The Bianchi identity allows us to solve 
Eq.~(\ref{Ein}) % Einstein equation 
%of  $F_{\mu\nu ab}$ 
for $\tilde{F}_{\mu\nu a b}$ leads to the constraint equation
in (\ref{cons}), which can be solved by
%$\nabla_{\mu}[\,\,^\ast\tilde{F}^{\mu\nu}+
%2\chi e^{-{\phi\over 2}-\sigma}\tilde{F}^{\mu\nu}-
%chi^2\,\,^\ast F^{\mu\nu}]=0$, by 
introducing $\bar{A}_\mu$ as
$\bar{F}_{\mu\nu}\equiv \partial_\mu \bar{A}_\nu 
-\partial_\nu \bar{A}_\mu 
=\tilde{F}_{\mu\nu}-2\chi\, 
e^{-{\phi\over 2}-\sigma}\,\,^\ast \tilde{F}_{\mu\nu}-\chi^2F_{\mu\nu}$. 
%Using this
% which is employed 
%With help of this relation,
The 10d equation %eq.(\ref{Hmnp}) 
for $H_{\mu\nu\lambda}$ is then solved by
\be
H_{\mu\nu\lambda}={3\over 2} 
e^{\phi-6\sigma}\, \epsilon_{\mu\nu\lambda}\,^\delta\,
\big(\,\pm A_\delta-\bar{A}_\delta-\nabla_\delta\psi\,\big)\,.
\ee
Using the Lagrange multiplier $\Lambda_{\mu\nu}$, 
%we may introduce the definition of $\bar{F}^{\mu\nu}$ as a constraint 
%on the action. 
we incorporate the above constraint into the Lagrangian.
After straightforward  calculation, we come to the %following 
dimensionally reduced effective Lagrangian in (\ref{effaction}).
%\bea
%16 \pi G_4 {\cal L}_{4}&=&
%R_4(g)-{1\over 2}(\nabla\phi)^2 - 6(\nabla\sigma)^2 - 
%{3\over 2} e^{-\phi-2\sigma}(\nabla\chi)^2 + 12e^{-4\sigma}  - {3\over 2} 
%e^{{3\phi\over 2}-5\sigma} - 6\chi^2e^{{\phi\over 2}-7\sigma} \nonumber\\
%&&- {9\over 2} e^{-{\phi\over 2}-9\sigma}(1+\chi^2)^2 - 
%e^{{3\phi\over 2}+3\sigma} \, F_{\mu\nu}F^{\mu\nu} - 
%18 e^{{\phi}-6\sigma}(A-\bar{A}-\nabla\psi)^2 -
% 3e^{-{\phi\over 2}-\sigma} \, \tilde{F}_{\mu\nu}\tilde{F}^{\mu\nu} 
% \nonumber\\
%&&+ 6e^{-{\phi\over 2}-\sigma}\Lambda^{\mu\nu}
%(\tilde{F}_{\mu\nu}-2\chi e^{-{\phi\over 2}-\sigma}
%\ast \tilde{F}_{\mu\nu}-\chi^2F_{\mu\nu}-
%\bar{F}_{\mu\nu}) - 4\chi^3 F_{\mu\nu}\ast F^{\mu\nu} \nonumber\\
%&&+ 6\chi \tilde{F}_{\mu\nu}(\ast F^{\mu\nu}+2\chi
% e^{-{\phi\over 2}-\sigma}F^{\mu\nu}) + 
%6\chi e^{-\phi-2\sigma}\tilde{F}_{\mu\nu}\ast \tilde{F}^{\mu\nu} \,.
%\eea
This Lagrangian possesses two kinds of auxiliary 
fields, $\tilde{F}_{\mu\nu}$ and $\Lambda_{\mu\nu}$. 
Variation with respect to $\tilde{F}_{\mu\nu}$ 
%endows us 
%with $\Lambda_{\mu\nu}$ in terms of other fields,
leads to
\be
\Lambda_{\mu\nu}=\tilde{F}_{\mu\nu}-\chi\, e^{{\phi\over 2}+\sigma}
\,\,^\ast F_{\mu\nu}\,,
\ee
while $\Lambda_{\mu\nu}$ variation gives us 
the constraint (\ref{cons}).
%By varying the Lagrangian 
The variations 
with respect to $\bar{A}_\nu$ and ${A}_\nu$ lead to%one can obtain the 
%equation of motion,
\be
\nabla_{\mu} \,\big(\,e^{-{\phi\over 2}-\sigma}\Lambda^{\mu\nu}\,\big)+
3e^{{\phi}-6\sigma}\,\big(\,\pm A^\nu-\bar{A}^\nu-\nabla^\nu\psi\,\big)=0\,,
\ee
and %its derivative with respect to $A_\nu$ gives
\be
\nabla_{\mu} \,\big(\,e^{{3\phi\over 2}+3\sigma}\,F^{\mu\nu})-3\nabla_\mu\chi
 \,\,^\ast\tilde{F}^{\mu\nu}-9\,e^{{\phi}-
6\sigma}\,(\pm 1+\chi^2)\,\big(\,\pm A^\nu-\bar{A}^\nu-\nabla^\nu\psi\,
\big)=0\,,
\ee
respectively.
The equations of motion for $\phi$ and $\sigma$ following from the 
effective
Lagrangian
%and $\chi$ 
%is 
%straightforward. They are
read
\bea
\nabla^2\phi &=& - {3\over 2} e^{-\phi-2\sigma}\,\big(\nabla\chi\big)^2 + 
{9\over 4}\, e^{{3\phi\over 2}-5\sigma}
 +3\chi^2\,e^{{\phi\over 2}-7\sigma} - 
{9\over 4}\, e^{-{\phi\over 2}-9\sigma}\,\big(\pm 1+\chi^2\big)^2 \nonumber\\
&&+ {3\over 2}\,e^{{3\phi\over 2}+3\sigma} \, F_{\mu\nu}\,F^{\mu\nu} + 
18 \,e^{{\phi}-6\sigma}\,\big(\,\pm A-
\bar{A}-\nabla\psi\,\big)^2 - {3\over 2}\,e^{-{\phi\over 2}-
\sigma} \, \tilde{F}_{\mu\nu}\,\tilde{F}^{\mu\nu} \,,\\
4\nabla^2\sigma &=& - e^{-\phi-2\sigma}\,\big(\nabla\chi\big)^2 + 
16\,e^{-4\sigma}  
- {5\over 2}\, e^{{3\phi\over 2}-5\sigma} -
 14\,\chi^2\,e^{{\phi\over 2}-7\sigma} -
 {27\over 2}\, e^{-{\phi\over 2}-9\sigma}\,\big(\pm 1+\chi^2\big)^2 \nonumber\\
&& + e^{{3\phi\over 2}+3\sigma} \, F_{\mu\nu}\,F^{\mu\nu} - 
36 e^{{\phi}-6\sigma}\,\big(\pm A-\bar{A}-\nabla\psi\big)^2 - 
e^{-{\phi\over 2}-\sigma} \, \tilde{F}_{\mu\nu}\,\tilde{F}^{\mu\nu} \,,
%\\
%\nabla_\mu(e^{-\phi-2\sigma}\nabla^\mu\chi)
% &=& 4\chi e^{{\phi\over 2}-7\sigma} + 
%6e^{-{\phi\over 2}-9\sigma}\chi(1+\chi^2) - 2 F_{\mu\nu}\ast 
%\tilde{F}^{\mu\nu} + 2e^{-\phi-2\sigma} \, 
%\tilde{F}_{\mu\nu}\ast\tilde{F}^{\mu\nu}\,.
\eea
which are consistent with the 10d equations.
%Interestingly, the terms in ${\cal L}_c$
%the last two lines of the effective Lagrangian 
%do not contribute to the equations for $\phi$ and $\sigma$, while 
%they do for $\chi$.
%One can also obtain the following Einstein equation 
From the metric variation, one has
\bea
R_{\mu\nu}&=&{1\over 2}\nabla_\mu\phi\,\nabla_\nu\phi + 
{6}\nabla_\mu\sigma\nabla_\nu\sigma + 
{3\over 2}e^{-\phi-2\sigma}\nabla_\mu\chi\,\nabla_\nu\chi
+2 e^{{3\phi\over 2}+3\sigma} \, \big(F_{\mu\alpha}F_\nu^{~\alpha}-
{1\over 4}g_{\mu\nu}F_{\alpha\beta}F^{\alpha\beta}\big)\nonumber\\
&+&
6e^{-{\phi\over 2}-\sigma} \, \big(\tilde{F}_{\mu\alpha}
\tilde{F}_\nu^{~\alpha}-{1\over 4}
g_{\mu\nu}\tilde{F}_{\alpha\beta}\tilde{F}^{\alpha\beta}\big)
 + 
18 e^{{\phi}-6\sigma}(\pm A_\mu-\bar{A}_\mu-\nabla_\mu\psi)\,
(\pm A_\nu-\bar{A}_\nu-\nabla_\nu\psi) 
 \nonumber\\
&-& {1\over 2}g_{\mu\nu}\Big[\,12e^{-4\sigma} 
 - {3\over 2} e^{{3\phi\over 2}-5\sigma} -
 6\chi^2\,e^{{\phi\over 2}-7\sigma}
- {9\over 2} e^{-{\phi\over 2}-9\sigma}\,(\pm 1+\chi^2)^2\,\Big] 
\,,
\eea
which is consistent with the 10d Einstein equation in (\ref{eq24}).
Note that one needs 
to impose $\tilde{F}_{\mu\nu}$ constraint 
before taking the metric variation.
Namely, one has to
 plug the constraint,  $\Lambda^{\mu\nu}
=\tilde{F}^{\mu\nu}-\chi\, e^{{\phi\over 2}+\sigma}
\,\,^\ast F^{\mu\nu}\!,\,$ into 
the action because 
the metric variation of
$\sqrt{-g}\,\Lambda^{\mu\nu}$ which involves
both $\sqrt{-g}\,\tilde{F}^{\mu\nu}$ and 
$\sqrt{-g}\,\epsilon^{\mu\nu\rho\sigma} F_{\rho\sigma}$ 
is not well defined before imposing 
constraint.

%%%%%%%%%%%%%%%%%%%%%%%%%%%%%%%%%%%%%%%%%%%%%%%%%%%%%%%%%%%%%%%%%%%%%%%%%%%

%%%%%%%%%%%%%%%%%%%%%%
%%%%%%%%%%%%%%%%%%%%%%%%%%%%%%%%%%%%%%%%%%%%%%%%%%%%%%%%%%%%%%%%%%%%%%%%%%%


\begin{thebibliography}{999}


%\cite{Maldacena:1997re}
\bibitem{Maldacena:1997re}
  J.~M.~Maldacena,
  ``The large N limit of superconformal field theories and supergravity,''
  Adv.\ Theor.\ Math.\ Phys.\  {\bf 2} (1998) 231
  [Int.\ J.\ Theor.\ Phys.\  {\bf 38} (1999) 1113]
  [arXiv:hep-th/9711200].
  %%CITATION = IJTPB,38,1113;%%

%\cite{Bak:2003jk}
\bibitem{Bak:2003jk}
  D.~Bak, M.~Gutperle and S.~Hirano,
  ``A dilatonic deformation of AdS(5) and its field theory dual,''
  JHEP {\bf 0305}, 072 (2003)
  [arXiv:hep-th/0304129].
  %%CITATION = JHEPA,0305,072;%%

%\cite{Clark:2004sb}
\bibitem{Clark:2004sb}
%\cite{Papadimitriou:2004rz}
%\bibitem{Papadimitriou:2004rz}
  I.~Papadimitriou and K.~Skenderis,
  ``Correlation functions in holographic RG flows,''
  JHEP {\bf 0410}, 075 (2004)
  [arXiv:hep-th/0407071];
  %%CITATION = JHEPA,0410,075;%%  
A.~B.~Clark, D.~Z.~Freedman, A.~Karch and M.~Schnabl,
  ``The dual of Janus $((<:)<-->(:>))$ an interface CFT,''
  Phys.\ Rev.\  D {\bf 71}, 066003 (2005)
  [arXiv:hep-th/0407073].
  %%CITATION = PHRVA,D71,066003;%%


%\cite{Bak:2007fk}
\bibitem{Bak:2007fk}
  D.~Bak, A.~Karch and L.~G.~Yaffe,
  ``Debye screening in strongly coupled N=4 supersymmetric Yang-Mills plasma,''
  JHEP {\bf 0708}, 049 (2007)
  [arXiv:0705.0994 [hep-th]].
  %%CITATION = JHEPA,0708,049;%%

%\cite{Bak:2007qw}
\bibitem{Bak:2007qw}
  D.~Bak, M.~Gutperle and A.~Karch,
  ``Time dependent black holes and thermal equilibration,''
  JHEP {\bf 0712}, 034 (2007)
  [arXiv:0708.3691 [hep-th]];
  %%CITATION = JHEPA,0712,034;%%
%\cite{Bak:2007jm}
%\bibitem{Bak:2007jm}
  D.~Bak, M.~Gutperle and S.~Hirano,
  ``Three dimensional Janus and time-dependent black holes,''
  JHEP {\bf 0702}, 068 (2007)
  [arXiv:hep-th/0701108].
  %%CITATION = JHEPA,0702,068;%%

%\cite{Aharony:2008ug}
\bibitem{ABJM}
  O.~Aharony, O.~Bergman, D.~L.~Jafferis and J.~Maldacena,
  ``N=6 superconformal Chern-Simons-matter theories, M2-branes and their
  gravity duals,''
  JHEP {\bf 0810}, 091 (2008)
  [arXiv:0806.1218 [hep-th]].
  %%CITATION = JHEPA,0810,091;%%


%\cite{Minahan:2008hf}
\bibitem{Minahan:2008hf}
  J.~A.~Minahan and K.~Zarembo,
  ``The Bethe ansatz for superconformal Chern-Simons,''
  JHEP {\bf 0809}, 040 (2008)
  [arXiv:0806.3951 [hep-th]];
  %%CITATION = JHEPA,0809,040;%%
%\cite{Bak:2008cp}
%\bibitem{Bak:2008cp}
  D.~Bak and S.~J.~Rey,
  ``Integrable Spin Chain in Superconformal Chern-Simons Theory,''
  JHEP {\bf 0810}, 053 (2008)
  [arXiv:0807.2063 [hep-th]].
  %%CITATION = JHEPA,0810,053;%%

%\cite{Bak:2009mq}
\bibitem{Bak:2009mq}
  D.~Bak, H.~Min and S.~J.~Rey,
  ``Generalized Dynamical Spin Chain and 4-Loop Integrability in N=6
  Superconformal Chern-Simons Theory,''
  Nucl.\ Phys.\  B {\bf 827}, 381 (2010)
  [arXiv:0904.4677 [hep-th]];
  %%CITATION = NUPHA,B827,381;%%
%\cite{Bak:2009tq}
%\bibitem{Bak:2009tq}
  D.~Bak, H.~Min and S.~J.~Rey,
  ``Integrability of N=6 Chern-Simons Theory at Six Loops and Beyond,''
  arXiv:0911.0689 [hep-th].
  %%CITATION = ARXIV:0911.0689;%%

%\cite{Gauntlett:2009zw}
\bibitem{Gauntlett:2009zw}
  J.~P.~Gauntlett, S.~Kim, O.~Varela and D.~Waldram,
  ``Consistent supersymmetric Kaluza--Klein truncations with massive modes,''
  JHEP {\bf 0904}, 102 (2009)
  [arXiv:0901.0676 [hep-th]].
  %%CITATION = JHEPA,0904,102;%%


%\cite{Hartnoll:2008vx}
\bibitem{Hartnoll:2008vx}
  S.~A.~Hartnoll, C.~P.~Herzog and G.~T.~Horowitz,
  ``Building a Holographic Superconductor,''
  Phys.\ Rev.\ Lett.\  {\bf 101}, 031601 (2008)
  [arXiv:0803.3295 [hep-th]];
  %%CITATION = PRLTA,101,031601;%%
%\cite{Hartnoll:2008kx}
%\bibitem{Hartnoll:2008kx}
  S.~A.~Hartnoll, C.~P.~Herzog and G.~T.~Horowitz,
  ``Holographic Superconductors,''
  JHEP {\bf 0812}, 015 (2008)
  [arXiv:0810.1563 [hep-th]].
  %%CITATION = JHEPA,0812,015;%%

%\cite{Basu:2008st}
\bibitem{Basu:2008st}
  P.~Basu, A.~Mukherjee and H.~H.~Shieh,
  ``Supercurrent: Vector Hair for an AdS Black Hole,''
  Phys.\ Rev.\  D {\bf 79}, 045010 (2009)
  [arXiv:0809.4494 [hep-th]].
  %%CITATION = PHRVA,D79,045010;%%

%\cite{Duff:1984hn}
\bibitem{Duff:1984hn}
  M.~J.~Duff, B.~E.~W.~Nilsson, C.~N.~Pope and N.~P.~Warner,
  ``On The Consistency Of The Kaluza-Klein Ansatz,''
  Phys.\ Lett.\  B {\bf 149}, 90 (1984).
  %%CITATION = PHLTA,B149,90;%%



%\cite{Nilsson:1984bj}
\bibitem{Nilsson:1984bj}
  B.~E.~W.~Nilsson and C.~N.~Pope,
  ``Hopf Fibration Of Eleven-Dimensional Supergravity,''
  Class.\ Quant.\ Grav.\  {\bf 1}, 499 (1984).
  %%CITATION = CQGRD,1,499;%%

%\cite{Casher:1984ym}
\bibitem{Casher:1984ym}
  A.~Casher, F.~Englert, H.~Nicolai and M.~Rooman,
  ``The Mass Spectrum Of Supergravity On The Round Seven Sphere,''
  Nucl.\ Phys.\  B {\bf 243}, 173 (1984).
  %%CITATION = NUPHA,B243,173;%%


%\cite{Cvetic:2000yp}
\bibitem{Cvetic:2000yp}
  M.~Cvetic, H.~Lu and C.~N.~Pope,
  ``Consistent warped-space Kaluza-Klein reductions, half-maximal gauged
  supergravities and CP(n) constructions,''
  Nucl.\ Phys.\  B {\bf 597}, 172 (2001)
  [arXiv:hep-th/0007109].
  %%CITATION = NUPHA,B597,172;%%

%\cite{Halyo:1998pn}
\bibitem{Halyo:1998pn}
  E.~Halyo,
  ``Supergravity on AdS(5/4) x Hopf fibrations and conformal field  theories,''
  Mod.\ Phys.\ Lett.\  A {\bf 15}, 397 (2000)
  [arXiv:hep-th/9803193].
  %%CITATION = MPLAE,A15,397;%%


%\cite{Balasubramanian:1998de}
\bibitem{Balasubramanian:1998de}
  V.~Balasubramanian, P.~Kraus, A.~E.~Lawrence and S.~P.~Trivedi,
  ``Holographic probes of anti-de Sitter space-times,''
  Phys.\ Rev.\  D {\bf 59}, 104021 (1999)
  [arXiv:hep-th/9808017].
  %%CITATION = PHRVA,D59,104021;%%

%\cite{Gauntlett:2009dn}
\bibitem{Gauntlett:2009dn}
  J.~P.~Gauntlett, J.~Sonner and T.~Wiseman,
  ``Holographic superconductivity in M-Theory,''
  Phys.\ Rev.\ Lett.\  {\bf 103}, 151601 (2009)
  [arXiv:0907.3796 [hep-th]];
  %%CITATION = PRLTA,103,151601;%%
%\cite{Gauntlett:2009bh}
%\bibitem{Gauntlett:2009bh}
  J.~Gauntlett, J.~Sonner and T.~Wiseman,
  ``Quantum Criticality and Holographic Superconductors in M-theory,''
  arXiv:0912.0512 [hep-th].
  %%CITATION = ARXIV:0912.0512;%%


%\cite{Chamblin:1999tk}
\bibitem{Chamblin:1999tk}
  A.~Chamblin, R.~Emparan, C.~V.~Johnson and R.~C.~Myers,
  ``Charged AdS black holes and catastrophic holography,''
  Phys.\ Rev.\  D {\bf 60}, 064018 (1999)
  [arXiv:hep-th/9902170].
 %%CITATION = PHLTA,B495,251;%%

%\cite{Bak:2006dn}
\bibitem{Bak:2006dn}
  D.~Bak and R.~A.~Janik,
  ``From static to evolving geometries: R-charged hydrodynamics from
  supergravity,''
  Phys.\ Lett.\  B {\bf 645}, 303 (2007)
  [arXiv:hep-th/0611304].
  %%CITATION = PHLTA,B645,303;%%

%\cite{Fujita:2009kw}
\bibitem{Fujita:2009kw}
  M.~Fujita, W.~Li, S.~Ryu and T.~Takayanagi,
  ``Fractional Quantum Hall Effect via Holography: Chern-Simons, Edge States,
  and Hierarchy,''
  JHEP {\bf 0906}, 066 (2009)
  [arXiv:0901.0924 [hep-th]];
  %%CITATION = JHEPA,0906,066;%%
%\cite{Hikida:2009tp}
%\bibitem{Hikida:2009tp}
  Y.~Hikida, W.~Li and T.~Takayanagi,
  ``ABJM with Flavors and FQHE,''
  JHEP {\bf 0907}, 065 (2009)
  [arXiv:0903.2194 [hep-th]].
  %%CITATION = JHEPA,0907,065;%%


%\cite{Rey:2008zz}
\bibitem{Rey:2008zz}
  S.~J.~Rey,
  ``String Theory On Thin Semiconductors: Holographic Realization Of Fermi
  Points And Surfaces,''
  Prog.\ Theor.\ Phys.\ Suppl.\  {\bf 177}, 128 (2009)
  [arXiv:0911.5295 [hep-th]].
  %%CITATION = PTPSA,177,128;%%



%\cite{Bak:2009kz}
\bibitem{Bak:2009kz}
  D.~Bak and S.~J.~Rey,
  ``Composite Fermion Metals from Dyon Black Holes and S-Duality,''
  arXiv:0912.0939 [hep-th].
  %%CITATION = ARXIV:0912.0939;%%


%\cite{Bak:2010qb}
\bibitem{kazem}
  D.~Bak, K.~B.~Fadafan and H.~Min,
  ``Static Length Scales of N=6 Chern-Simons Plasma,''
  arXiv:1003.5227 [hep-th].
  %%CITATION = ARXIV:1003.5227;%%




\end{thebibliography}
\end{document}